# Machining of complex shaped parts with guidance curves

**Laurent Tapie** [*,a,b] , **Bernardin Mawussi** [a,b], **Walter Rubio**[c], **Benoît Furet**[d]

[*] Corresponding author, laurent.tapie@univ-paris13.fr, Tel: +331 58 07 67 25, Fax: +33 1 58 07 67 25

[a]Unité de Recherches Biomatériaux Innovants et Interfaces – URB2i ; 1 Rue Maurice Arnoux; 92120 Montrouge, France

[b] IUT de Saint Denis, Université Paris 13; Place du 8 mai 1945 ; 93206 Saint Denis Cedex, France

[c] ICA (Institut Clement Ader) ; 135 Avenue de Rangueil ; 31077 Toulouse, France

[d] IRCyN, Ecole Centrale de Nantes ; 1 rue de la Noë ; 44321 Nantes Cedex 3, France

*Nowadays, High Speed Machining (HSM) is usually used for production of hardened material parts with complex shapes such as dies and moulds. In such parts, toolpaths generated for bottom machining feature with the conventional parallel plane strategy induced many feed rate reductions especially when boundaries of the feature have a lot of curvatures and are not parallel. Several machining experiments on hardened material lead to the conclusion that a tool path implying stable cutting conditions might guarantee a better part surface integrity. To ensure this stability, the shape machined must be decomposed when conventional strategies are not suitable. In this paper, an experimental approachbased on High Speed Performance simulationis conducted on a master bottom machining feature in order to highlight the influence of the curvatures towards a suitable decomposition of machining area. The decomposition is achieved through the construction of intermediate curves between the closed boundaries of the feature. These intermediate curves are used as guidance curve for the toolpaths generation with an alternative machining strategy called "guidance curve strategy". For the construction of intermediate curves, key parametersreflecting the influence of their proximity with each closed boundary and the influence of the curvatures of this latterare introduced. Based on the results, a method for defining guidance curves in 4 steps is proposed.*

## 1 Introduction

High Speed Machining (HSM) production of hardened material parts with complex shapes such as dies and mouldsis an economical process. Indeed part production is faster for better quality and longer fatigue life for the parts, butthecutting process induces the use of specific machining technics and technologies[1][2].Moreover tasks carried out during the preparationstep are mainly based on Computer-Aided Machining (CAM) software thatprovide several machining strategies[3][4]. Selection of these strategies is a very complex task. Indeed, several parameters in different fields (quality, cost, geometry ...) should be consideredduringmachining strategy selection and toolpaths generation. Besides commercial CAM solutions with relevant assistance modules for machining strategies selection are scant[5][6].

HSM cutting characteristicsstand forone of the main topics difficult to integrate in CAM software. Unfortunately, HSM cutting phenomenonhas a direct influence on part surface integrity[7].Importance of integrating the characteristics and impacts of the HSM in tool paths generation istherefore criticalfor complex shaped parts[8]. The correlation between part's topology (shape, size and local curvature of mould and die features for example), cutting tool types, cutting parameters (depths of cut, cutting speed, feedrate) and machining strategies (machining direction, sweeping direction)has been underlinedin several works[9][10]. Some studies on surface integrity after hard milling of mould and die (hardened steel)highlight the links between part surface residual stress, effective cutting speed,effectivefeedrate and contact area of tool/part[11][12][13].Parameters defining machining strategies are also relevant factors for obtaining surface integrity. The residual normal stresses along feed and stepover directions increased with the radial depth of cut and feed rate increase[14].For the works presented in this paper only the topology of the parts, the cutting conditions and the machining strategy were retained in the experimental study.

A tool path implying stable cutting conditionsmight guarantee a better part surface integrity.To ensure this stability, the shape machined must be decomposed when conventional strategies are not suitable. In this paper, experiments are carried out in order to analyse the influence of the decomposition. Machining performance simulated during the decomposition processis evaluated towards feed rate reductions (main criteria), machining time and tool path lengths. A bottom feature extracted from complex forging dies is the shape considered because it cannot be machined easily with conventional machining strategies such as parallel planes.First, the experimental setups including a module for High Speed Performance evaluation are presented. Then, the decomposition method is detailed through the construction of the intermediate curves and machining areas.Finally, the analysis of the results leads to the proposal of a first definition of guidance curves.

## 2 Experimental setup

### 2.1 Experimental machining feature

The machining featurestudied in this work is atypical forging die machining feature called "bottom feature". According to the definition developed in previous works[15][16], a bottom feature is generally in contact with two flanks in order to form a cavity(Fig. 1(a), (b)). In general, such cavitycomes out in other cavities on each openedboundary. Due toitsown basic geometryand the topology of its relationship with flanks, a bottom feature may induce more or less complex machining tool paths.

The master bottom feature presentedin this paper(Fig. 1(c))is in contact with flanks along two edges very curved and non-parallel in order to study their influence on the machining tool paths. Besides, the topology of this master bottom featureis much curved.

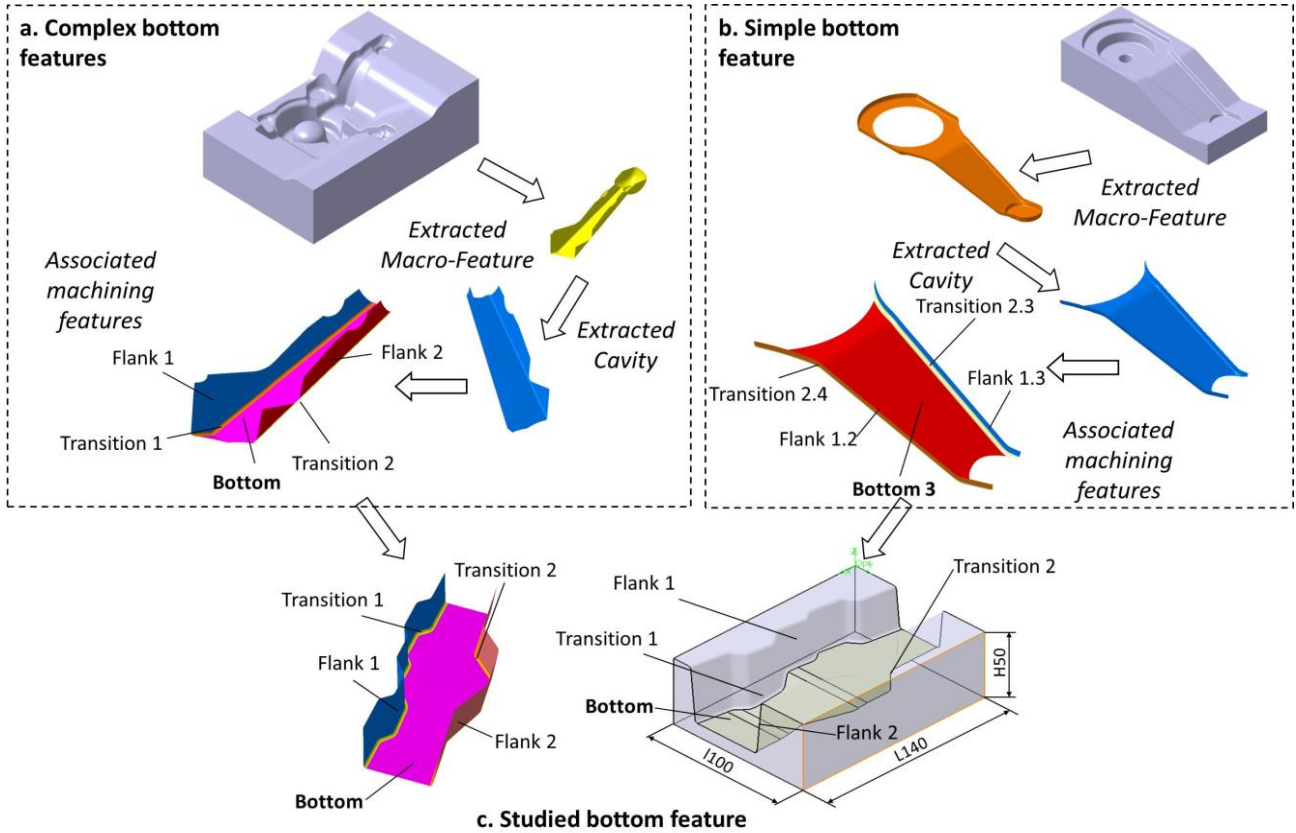

Fig. 1    Bottom featuresand experimental machining feature

### 2.2 Guidance curve machining strategy

Traditionally, a machining assistant uses standard 3-axes parallel planes strategy widely available in CAM software for the machining of bottom features[17][18]. A key characteristic of this machining strategy at the finishing stage is that the parallel planes contain the tool axis, which also defines the machining feed direction (Fig. 2(a)). The basic direction of the parallel planes is defined from the opened sides of the bottom feature. This basic direction can be adjusted according to topological information defined in the bottom machining feature model[15][16].

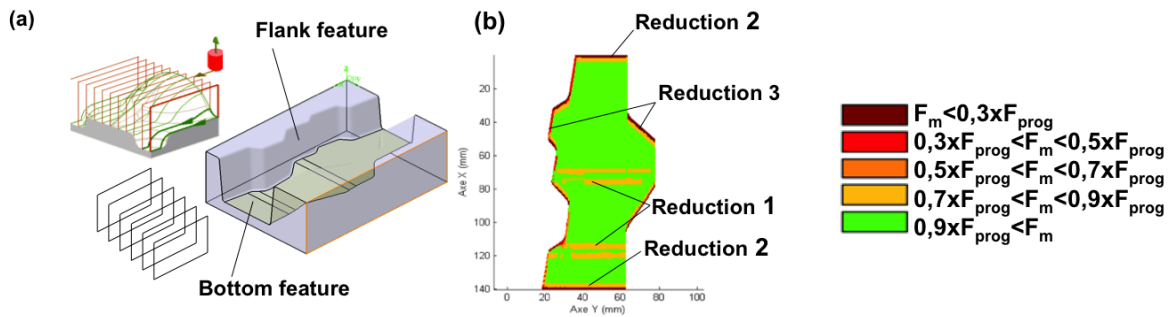

Fig. 2    Parallel planes tool path analysis

Analysis and simulation of the tool paths generated for the feature presented in Fig. 2show that the feed rate set point is not reached on the whole surface of the bottom feature (Fig. 2(b)). Three types of feed rate reductions can be distinguished. Feed rate reductions induced by small curvature radii defined in the CAD model of the feature (type 1). This type of feed rate reduction can be deleted only if the geometry of the feature is modified in the CAD model. Feed rate reductions closed to opened boundaries of the bottom feature (type 2) can be avoided if the tool paths are extended over the boundaries. Flank features induced feed rate reductions (type 3) because their intersections with the bottom feature are not straight and parallel lines. This type of feed rate reductions, which mainly correspond to tool entrance/exit, can be partially avoided by choosing another machining strategy which is more closed to the topology of the bottom feature. Machining strategies must be chosen appropriately in order to avoid feed rate reductions and tool entrances/exits because they have an impact on roughness and the part surface integrity (local crack, local hardness variations). Parallel plane strategy used for the example presented in Figure 1 is relevant for the central area of the bottom feature whereasalong the limits of flank featuresit is not relevant according to feed rate analysis.

An alternative machining strategy should be chosen or developed for processing these constraints induced by the flank features, which have a particular topology.In this paper, an alternative strategy, called "guidance curve" strategy, is studied. This work is carried out on tool paths computation using the guidance curves strategy implemented on the CAM software CATIA® V5. The implementation of this strategy requires first the definition of two guidance curves (Fig. 3) and then the generation of tool paths by morphing between these two guides curves.

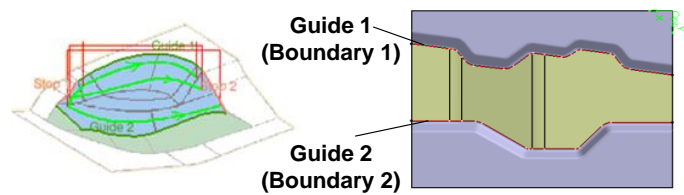

Fig. 3    Guidance curves strategy

## 2.3  High Speed Performance evaluation

Machining time and geometrical allowance of parts are the main components of High Speed Performance (HSP). In this work HSP is evaluated by the tool paths analysis: ISO block extracted from the machining program sent to the machine tool are analysed according to its length and the mean value of the simulated feed rate reached compared with the set point. The HSP is evaluated using the "performance viewer" module (Fig. 4.a.). The feed rate simulation principle is detailed on Fig. 4.b. The feed rate simulator implementation is based on a controlled jerk kinematical model and a discontinuity crossing kinematical model of the couple machine tool & NCU[19][20]. The performance viewer outputs are a ISO block length classification (histogram), a ISO block simulated feed rate classification (histogram), the simulated machining time and a map of the simulated feed rate on the whole toolpath.

For this work, the feed rate simulation is implemented with the parameters of MIKRON UCP 710 machine tool associated to a Siemens 840D NCU. The characteristics of the NCU are: X-Y-Z-axis maximal feed rate is 30 m/min, X-axis maximal acceleration is 2.5 m/s$^2$, Y-axis maximal acceleration is 3m/s$^2$, Z-axis maximal feed rate is 2 m/s$^2$, X-Y –axis maximal jerk is 5 m/s$^3$ and Z-axis maximal jerk is 50 m/s$^3$. The following cutting conditions are adopted: 4mm ball end mill cutting tool, machining deviation fixed to 0.01mm, cusp height fixed to 0.01mm, and the machining area sweeping set to upward. Tool path feed rate simulations are done for 2, 4 and 6 m/min set point.

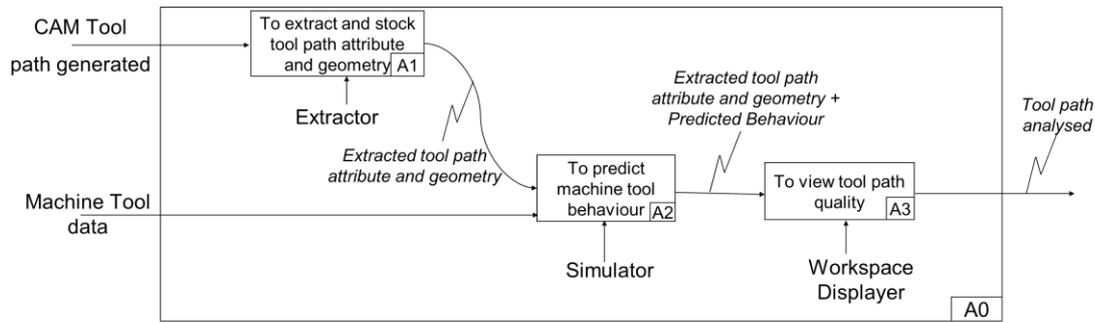

a. Performance viewer: global structure

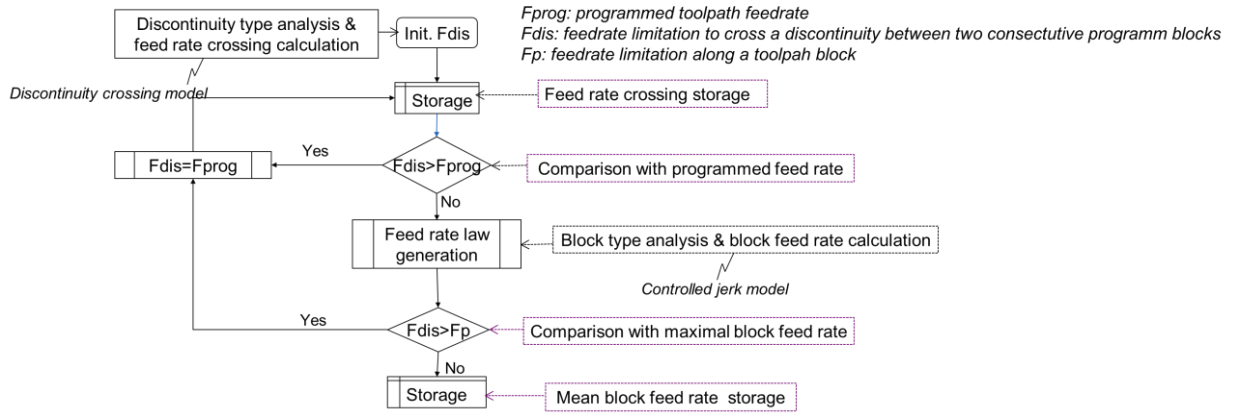

b. Performance viewer: simulator structure (A2)

Fig. 4    Performance Viewer

## 3    Construction of machining area

The whole area between the two boundaries of the bottom feature is the first machining area to be evaluated. This initial area is the result of the design of the shape and the different curvatures of the boundaries will spread on tool paths. The insertion of intermediate curves between the two boundaries can limit this propagation while using it as a basis for analysing the influence of curvatures. The association of the intermediate curves and the two boundaries results in the decomposition of the initial zone into multiple machining areas.

### 3.1    Single machining area

A preliminary decomposition in a single machining area defined between the guides 1 and 2 of the master 'bottom feature' is experimented (Fig. 3).

Analysis of the tool paths generated shows a very limited number (less than 1% of the total number) of short length machining sequences (Fig. 5(b) compared to parallel planes strategy (about 40% of the total number of trajectories, Fig. 5 (a)). Yet, the HSP cannot really be considered better since the number of entrance/exit in part material is significantly reduced, but the machining time is higher (579s.) compared to the for parallel planes strategy (418s).

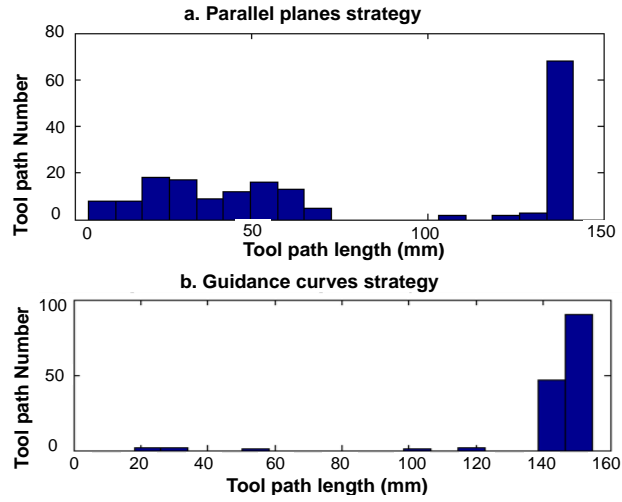

Fig. 5      Trajectories length classification

Besides, significant feed rate reductions appear on the central area of the bottom feature (Fig. 6). These reductions derive from the combination of two topological constraints: the bottom feature topology (small radii of curvature) as it was underlined for the parallel planes strategy (reduction type 1) and the two guide's topology. Indeed, significant feed rate reductions are induced by spreading of small curvature radii of the two guides (red colour in Fig. 6). The guidance curve strategy removes tool entrance/exit paths along boundaries of the bottom feature but it spreads some local cutting problems on the central area.

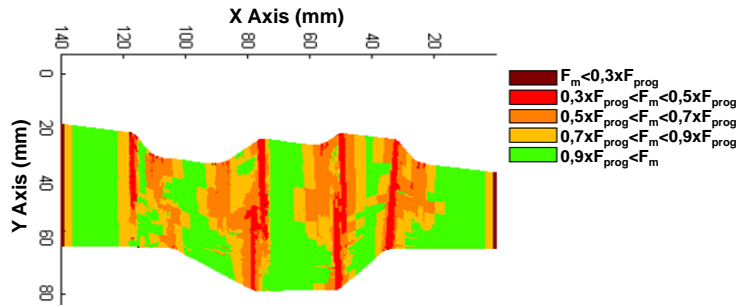

Fig. 6      Feed rate simulation for guidance curves strategy

To reduce these spreading influences, approach adopted in this work consists in decomposing the bottom feature area in several sub-areas. Limits, position and size of each sub-area should be defined in order to combine advantages of guidance curves and parallel planes strategies. After the decomposition, each new machining area is a portion of the initial machining feature limited by two curves from which one at least is different from the boundaries of the bottom feature. Several alternative guidance curves are experimented in order to define the machining area limits, position and size.

*3.2 Construction of intermediate curves*

Curves limiting the sub-areas are generated between the two boundaries of the bottom feature in an iterative way. To initialise the iteration process, the curve $C_{Int\_1}$ is generated between the two boundaries $B_1$ and $B_2$ which are the boundaries of the bottom feature. This initialisation step principle is illustrated in (Fig. 7.a.). The two initial boundaries $B_1$ and $B_2$ are discretized. The discretization is based on the generation of parallel planes spacing by a step P ($P_{Lm}$ in Fig. 7.a.). Each plane direction is perpendicular to the machining direction of the bottom feature given by topological decomposition explained in [15][16]. In this work the machining direction was found parallel to the length "L140" in Fig. 1. Then the intersecting points, $Pt_{m1}$ and $Pt_{m2}$ in Fig. 7.a., are generated between the two boundaries $B_1$ and $B_2$. According to principle presented to the equation (1), for a given value for K, intermediates points $Pt_{m\_Int\_1}$ are generated in all planes $P_{Lm}$ between $Pt_{m1}$ and $Pt_{m2}$. K is a factor introduced in order to represent the topological influence of each boundary $B_1$ or $B_2$ on the generation of intermediate curves $C_{Int\_1}$ (Fig. 7.a.).

$$\overrightarrow{Pt_{m1}Pt_{m\_Int\_1}} = K \times \overrightarrow{Pt_{m1}Pt_{m2}}, \quad K \in \,]0,1[ \qquad (1)$$

Finally the intermediate curve $C_{Int\_1}$ is created as a 5-degree B-spline curve controlled by points $Pt_{m\_Int\_1}$, 5-degree B-Spline model implemented in CATIA® V5.

Then, iteration process is based on the generalization of the initialization process for two curves $C_a$ and $C_b$, which are not intersected (Fig. 7.b.). Each point $Pt_{m\_Int\_j}$ defining the intermediate curve is computed using the following equation 2.

$$\overrightarrow{Pt_{ma}Pt_{m\_Int\_j}} = K \times \overrightarrow{Pt_{ma}Pt_{mb}}, \quad K \in\; ]0, 1[ \qquad (2)$$

Where:

$Pt_{ma}$ and $Pt_{mb}$ are intersection points of $P_{Lm}$ with the two curves $C_a$ and $C_b$.

$P_{Lm}$ is one of the discretization planes spacing by a step P and perpendicular to the machining direction of the bottom feature.

For a given value for K, points $Pt_{m\_Int\_j}$ are generated in all planes $P_{Lm}$. The intermediate curve $C_{Int\_j}$ is created as a 5-degree B-spline curve controlled by points $P_{tm\_Int\_j}$, 5-degree B-Spline model implemented in CATIA® V5.K is a factor introduced in order to represent the topological influence of each curve $C_a$ or $C_b$ on the generation of intermediate curves $C_{Int\_j}$.

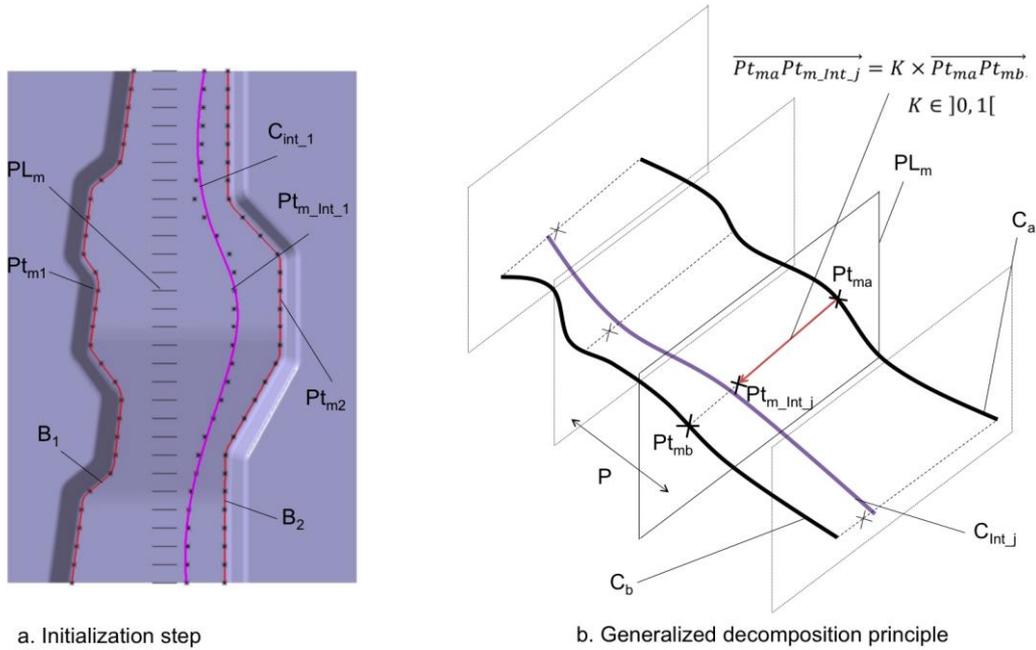

a. Initialization step

b. Generalized decomposition principle

Fig. 7      Intermediate curve building principle

Then, a net of intermediate curves is generated according to the principle presented below. Each curve $C_{Int\_j}$ (j>1) is generated between $C_{Int\_j-1}$ and $B_1$ or $B_2$. The iterative process stops when the current intermediate curve $C_{Int\_j}$ and $B_1$ or $B_2$ are intersecting, i.e. $C_{Int\_j} \cap B_1$ or $B_2 \neq \{\varnothing\}$. The last curve saved is $C_{Int\_j-1}$. Thus, the net of curves generated is {B1, $C_{Int\_1}$, ..., $C_{Int\_j-1}$, $B_2$}. For the example presented in Fig. 8 (a), the iterative process is applied from $B_1$ to $B_2$, $C_{Int\_10}$ intersects $B_2$ (Fig. 8 (b)), thus $C_{Int\_9}$ is the last curve generated in the net.

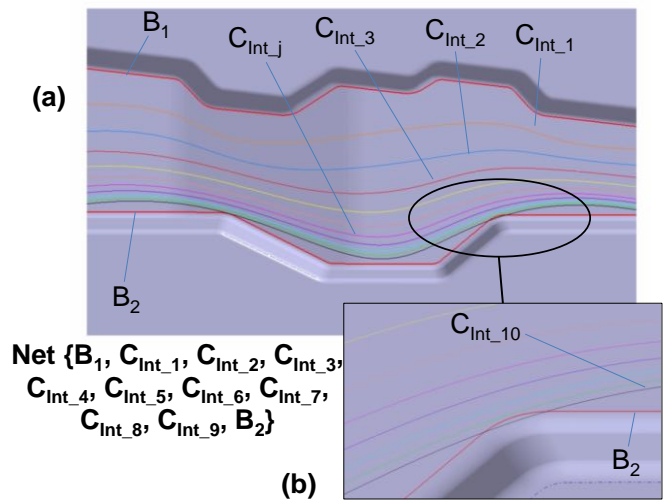

Fig. 8    Curves net building principle for K=0.25 and P=5mm

The Composed machining area is defined as the association of elementary machining areas covering the global machining area of the bottom feature. Each elementary machining area is limited by two consecutive curves of the net generated.

### 3.3 Construction of multiplemachining area

#### 3.3.1 Boundary directionmachining area

The first type of decompositionis carried out with two nets of intermediate curves: from boundary 1 ($B_1$) to boundary 2 ($B_2$) and from boundary 2 to boundary 1. For K=0.25 nine intermediate curves from $B_1$ to $B_2$ are generated and illustrated in Fig. 9. Based on these intermediate curves, the machining area between $B_1$ and $B_2$ is decomposed in an iterative way: $M_{A1} = \{B_1, C_{Int\_1}\} \cup \{C_{Int\_1}, B_2\}$, $M_{A2} = \{B_1, C_{Int\_1}\} \cup \{C_{Int\_1}, C_{Int\_2}\} \cup \{C_{Int\_2}, B_2\}$, and for j=3 to 9 $M_{Aj} = \{B_1, C_{Int\_1}\} \cup \{C_{Int\_1}, C_{Int\_2}\} \cup \ldots \cup \{C_{Int\_i-1}, C_{Int\_i}\} \cup \ldots \cup \{C_{Int\_j}, B_2\}$. Machining simulation of these nine decomposed areas allows analysing the spreading of boundary 1's topology constraints.

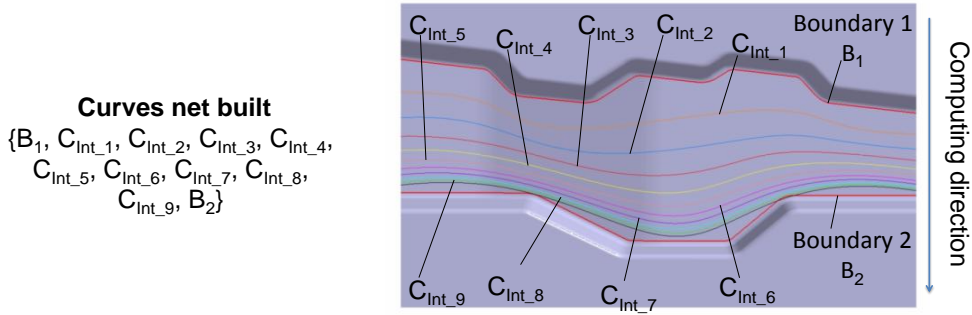

Fig. 9    Composed machining areas (K=0.25)

The spreading influence is analysed for K=0.75 and the decomposition process is carried out also in the second direction from $B_2$ to $B_1$. All the machining areas decompositiontype 1 experimented are presented in Table. 1.

| P | Curves net characteristics | | | Number of elementary machining areas |
|---|---|---|---|---|
| | Building direction | K | $C_{Int}$ number | |
| 5mm | Boundary 1 to Boundary 2 | 0.25 | 9 | 2,3, 4, 5, 6, 7, 8, 9, 10 |
| | | 0.75 | 1 | 2 |
| | Boundary 2 to Boundary 1 | 0.25 | 9 | 2,3, 4, 5, 6, 7, 8, 9, 10 |
| | | 0.75 | 1 | 2 |

Table. 1    Composed machining areas type 1

#### 3.3.2 Median curve machining area

The second type of decomposition is carried out with a pre-decomposition into two machining areas at the initialisation step. A median curve ($C_{med}$) is generated according to the intermediate curve principle with K=0.5. This value is chosen in order to expect the same influence of the two boundaries in each pre-decomposed machining area. Intermediate curves' net are generated in each two pre-decomposed area: in one hand from $B_1$ and $B_2$ to the median curve, and in the other hand from the median curve ($C_{med}$) to $B_1$ and $B_2$ (Fig. 10). The composed machining areas are built based on the same principle presented previously. For K=0.5, six elementary machining areas are generated, three areas from the median curve $C_{med}$ to $B_1$ and three other areas from $C_{med}$ to $B_2$ (illustrated in Fig. 10).

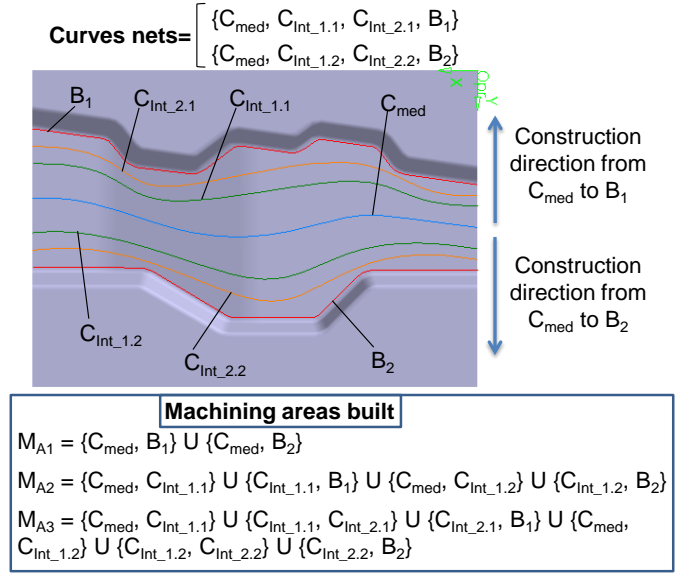

Fig. 10    Composed machining areas (K=0.5)

All the composed machining areas generated are presented in Table. 2. The number of intermediate curves and their position with respect to the boundaries are defined with: K=0.25, K=0.5 and K=0.75. Besides the number of control points used to generate intermediate curves is also defined through different step values: P=5mm, P=10mm and P=20mm between two consecutive control points.

| P | Curves net characteristics | | | Number of elementary machining areas |
|---|---|---|---|---|
| | Building direction | K | $C_{Int}$ number | |
| 5mm, 10mm, 20mm | Boundary 1 to $C_{med}$ and Boundary 2 to $C_{med}$ | 0.25 | 13 | 2, 4, 6, 8, 10, 12, 14 |
| | | 0.5 | 5 | 2, 4, 6 |
| | | 0.75 | 3 | 2, 4 |
| | $C_{med}$ to Boundary 1 and $C_{med}$ to Boundary 2 | 0.25 | 13 | 2, 4, 6, 8, 10, 12, 14 |
| | | 0.5 | 5 | 2, 4, 6 |
| | | 0.75 | 3 | 2, 4 |

Table. 2    Composed machining areas built on curves net with median

## 4 Results and discussion

### 4.1 Analysis of boundary direction machining area

For the machining area created with K=0.75 (illustrated in Fig. 11 (b)) according to the building direction Boundary 1 to Boundary 2, the machining time is equal than those of the single machining area (between boundaries 1 and 2 illustrated in Fig. 11 (a)) and toolpath lengths are higher than those of the single machining area. As might be expected, the decomposition of the bottom surface in two elementary machining areas (Fig. 11 (b)) involves some limiting feed rate reductions around the intermediate curve compared with the single machining area (Fig. 11 (a)). This observation is reinforced by the classification of blocks according to the feed rate. The number of short tool paths does not change but their feed rates remain as low as for the single machining area. The addition of intermediate curves seems relevant to limit only feed rate reductions.

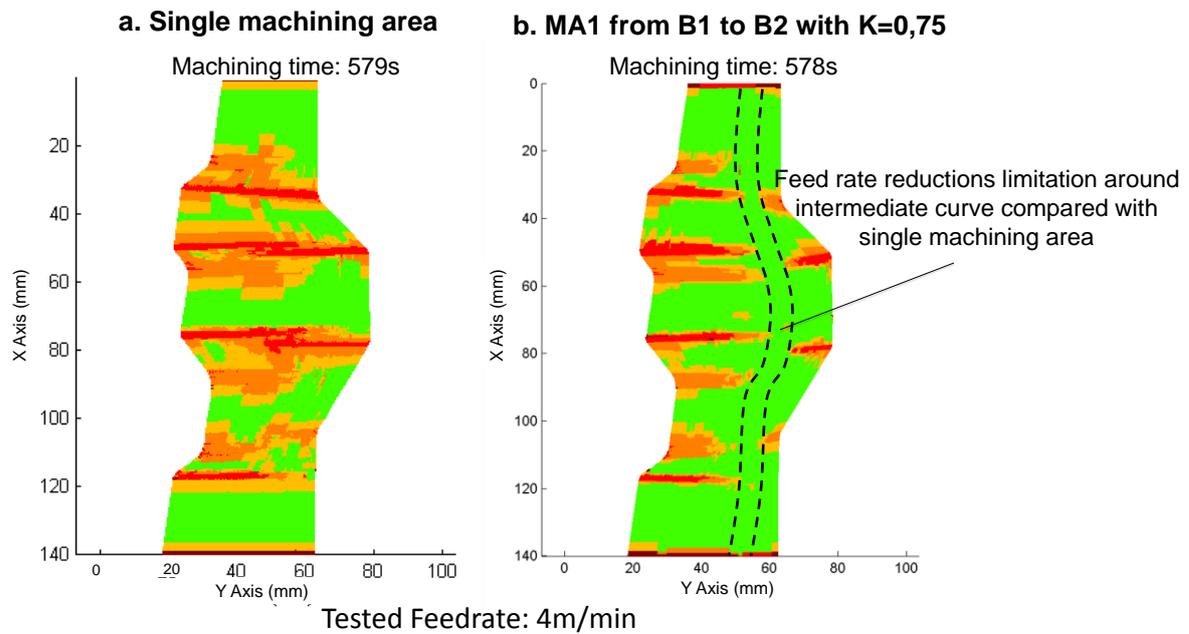

Fig. 11　　Feed rate simulations for single machining area and composed machining area MA1(K0,75)

When the initial area is decomposed in 2 elementary areas with K=0.25 ($M_{A1}$ equivalent to K=0.75 in the direction from boundary 2 to boundary 1), machining time, tool paths length and block's number in each feed rate class are equivalent. Some feed rate reductions are localized around the intermediate curve compared with the single machining area as it is previously observed.

For the composed machining areas obtained for K=0.25, machining time and tool path length remain higher than those of the single machining area. As might be expected, values of these two criteria increase with the number of elementary machining areas, i.e. the addition of intermediate curves implies increase of machining time and tool path length. Yet, addition of intermediate curves limits feed rate reductions compared with the single machining area. In this case, remaining feed rate reductions are reflected towards the second boundary in the construction direction of intermediate curves. Fig. 12 shows the comparison between machining areas $M_{A1}$ and $M_{A9}$. The reflection of feed rate reductions towards boundary 2 is clearly visible (area 2 in Fig.13 (a) and (b)) for the decomposition $M_{A9}$ (10 elementary machining areas). Feed rate reductions appearing in the first elementary machining area (area 1 between $B_1$ and $C_{Int1}$ in Fig.13 (a) and (b); require the construction of additional intermediate curves between $B_1$ and $C_{Int1}$ for the building direction $B_1$ to $B_2$. Improvement made by this construction of additional curves is observed when the construction direction changes from $B_2$ to $B_1$ (area 1 between $B_2$ and $C_{Int1}$ in Fig.13 (c)).

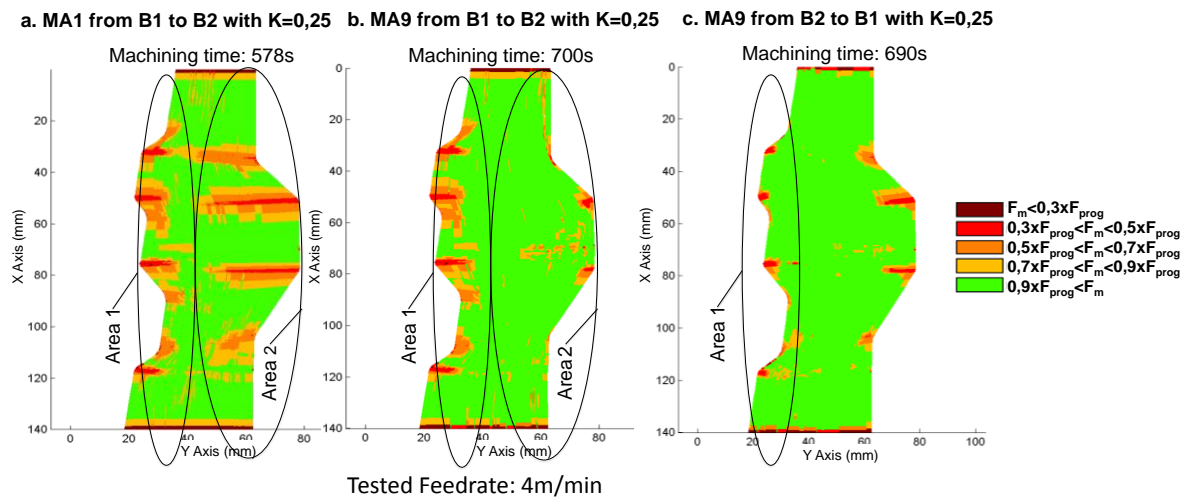

Fig. 12　　Feed rate simulations for composed machining areas $M_{A1}$ and $M_{A9}$

The generation of elementary areas by construction of intermediate curves between the two boundaries $B_1$ and $B_2$ reflects feed rate reductions towards the second boundary according to the construction direction. The first elementary area generated integrating the first boundary is to large to give the same result as that observed close to the second boundary. For this reason, the construction of intermediate curves in the two directions must be combined. This combination requires the division of the single machining area (between $B_1$ and $B_2$) into two areas with a main intermediate curve. The shape and the position of this main intermediate curve have to be defined.

The comparison between 2 elementary areas generated for K=0.25 ($M_{A1}$) or K=0.75 shows that the machining time and the number of feed rate reductions are not affected by the position of the intermediate curve with respect to the boundaries. Yet, curves limiting elementary machining areas will generate feed rate reductions.

Addition of several intermediate curves implies that machining time increases (until 20% for K=0.25) whereas feed rate reductions are very limited. This observation can be explained by the evolution of curvature radii (from small to soft radii) on the limits of the elementary machining areas. The number of intermediate curves might be maximized to limit feed rate reductions but this maximization will increase the machining time. The number of intermediate curves and their proximity to the boundaries must be analysed to find the better balance between feed rate reductions (parts geometrical and mechanical requirements) and machining time (productivity).

*4.2 Analysis of median curve machining area*

In the first step of these experiments, elementary areas are generated for a given value for K, by addition of intermediate curves from median curve to respectively $B_1$ and $B_2$. As it might be expected, feed rate reductions are reflected towards each boundary and the machining time increases with the number of intermediate curves.

When construction directions of intermediate curves are inverted (from each boundary to median curve), the machining areas created limit the feed rate reductions in the elementary areas closed to the boundaries. The number of elementary machining areas increases with the machining time but it does not affect feed rate reductions.

A first conclusion can be made: the construction of intermediate curves gives best results from the median curve to the each boundary for the limitation of feed rate reductions. The number of intermediate curves increases with the machining time as it was previously observed in the first experiments.

Feed rate simulations for a decomposition of the single area into 4 elementary areas are presented in Fig. 13. The comparison of these three simulations highlights a better limitation of feed rate reductions when elementary areas are positioned more close to the boundaries for K=0.75 (Fig. 13(c)). In these three cases, the machining time remains equivalent. The size of the first elementary machining area for K=0.75 (Fig. 13(c)) is greater than for K=0.5 (Fig. 13(b)), itself being greater than for K=0.25 (Fig. 13(a)).

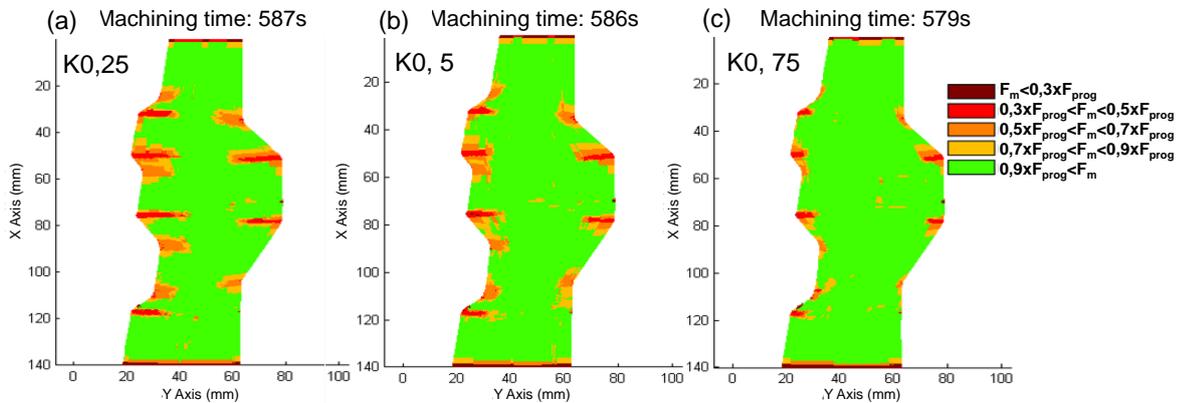

Fig. 13    Feed rate simulations for composed machining areas with 4 elementary areas

Feed rate simulation for the decomposition of the single machining area into 4 elementary areas for K=0.75 (Fig. 14 (a)) is compared to the one realized for the decomposition into 14 elementary areas for K=0.25 (Fig. 14 (b)). For each one of these simulations, several elementary machining areas are localized close to the boundaries. The number of elementary machining areas in the central area is greater for K=0.25 than for K=0.75. When consider feed rate reductions, results are equivalent for the two decompositions. Yet, the machining time is greater for K=0.25.

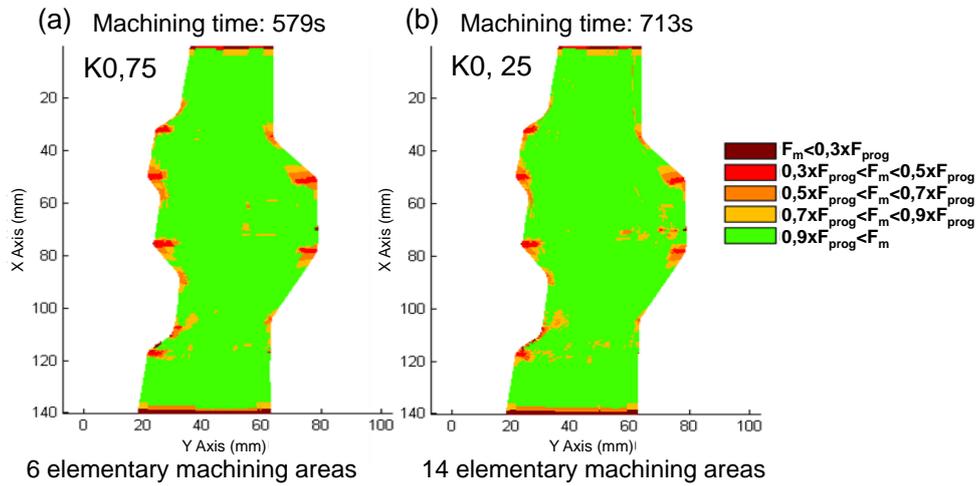

Fig. 14  Feed rate simulations for more than 4 elementary areas

A second conclusion can be made: to limit feed rate reductions for a same number of elementary machining areas and a same (or equivalent) machining time, it is necessary to localize elementary machining areas close to the boundaries. The number of elementary machining areas, i.e. the number of intermediate curves close to boundaries increases with the machining time without any significant change for feed rate reductions. According to experiments carried out, the decomposition of the single machining area into 4 elementary areas is the most relevant with a value for K close to 0.75.

As last conclusion concerning the intermediate curve smoothing, the variation of the step P does not have a significant influence on the results already observed. However, it is advisable to retain its smallest value than possible which can be set to 5 mm.

4.3  Definition of guidance curves

Analysis of the experiments performed shows that the appropriate solution for the 'bottom machining' feature proposed in this work is to decompose the single machining area of the bottom machining feature in 4 elementary areas generated with 3 intermediate curves. The first intermediate curve is the median curve ($C_{med}$) generated between the boundaries with K=0.5 according to the principle presented in this paper. The two others intermediate curves ($C_{Int\_1}$ and $C_{Int\_2}$) are generated with K=0.75 between the median curve ($C_{med}$) and each boundary ($B_1$ and $B_2$). Based on the experiments, a first definition of guidance curves can be proposed in four steps for a given cavity:

Step 1: Generate the median curve $C_{med}$ between the boundaries $B_1$ and $B_2$ with K=0.5. This median curve separates the initial machining area into two sub-areas.

Step 2: Determine the appropriate step P for smoothing intermediate curves. Indeed, the discretization of the two guidance curves with the initial step set to 5 mm can lead to very few points or no points when the radius of curvature is very small. The projection of the inflection points of the two guidance curves on a line parallel to the machining direction makes it possible to set the suitable value of the step or to adapt it along the machining direction.

Step 3: For each sub-area limited by two guidance curves which correspond to the median curve and one boundary $B_1$ or $B_2$, generate intermediate curves with K equal to 0.75, starting from the guidance curve which smallest radius of curvature is greatest. A line as a guidance curve is still considered as the starting curve and when the two guidance curves are lines, it is not necessary to generate intermediate curves.

Step 4: Generate and simulate tool paths for all machining areas created. The value of K can be increased or decreased slightly to see if there is improvement.

The first method proposed to define the guide curves provides assistance for machining specialists. The method will be generalized based on further experiments being conducted. This generalization will be communicated later.

5  Conclusion

The machining of complex shapes is often based on two conventional strategies: parallel plane and Z level. Tool paths generated with these conventional strategies for a bottom machining feature have many feed rate reductions especially when both boundaries have a lot of curvatures and are not parallel. Experimental approach presented in this paper is aimed to highlight the influence of the curvatures towards a suitable decomposition of the machining area limited by both boundaries.It also introduces one of theuncommon methods of implementation of the curve guidance machining strategy available in CAD/CAM software.

The bottom featureusually present in the cavities of complex shapediesis part of the experimental setup. Difficulties of generating tool paths often encountered with this machining feature were first analysed towards parallel planes strategy before the decomposition of the machining area. Feed rate reductions highlighted in the scope of the first analysis are completed by the high performance evaluation which is carried out by means of the computation of tool paths' length for each ISO block extracted from the machining program and sent to the machine tool. Other components of high performance as the machining time and the geometrical allowance are also evaluated but they are not presented because their variations are not significant.Application of the approach to other types of open machiningfeatures (non-closed boundary) is trivial when the two limit curves are identified and extracted.

The decomposition of the machining area is achieved through the construction of intermediate curves between the two boundaries. At first, a series of points, proportionally spaced from both boundaries are created before they are approximated to create each curve.The ratio of the distances between points and a givenboundary reflects the influence of its curvatureson the intermediate curve.The ratio of the distances is the first key parameter of the proposed approach.To prevent the spread of the influence of the curvatures of one boundary to the other boundary, the median curve was finally created. It shares the machining area in two portions before the decomposition.

Machining areas evaluated are constructed by combining intermediate curves created before. Results were analysed at several levels. First, it was shown that increasing the number of intermediate curves tends to limit the feedrate reductions while also increasing the machining time.Then, it was noted that the introduction of the median curve leads to a significant decrease in feed rate reductions when the intermediate curves are constructed from this curve to the boundaries.Finally, the observations show that the positioning of the machining areas (intermediate curves) closest to the boundaries gives better results and the variation of the curves smoothing step P has not a significant influence on the results. Based on the results, a method of defining guidance curves in 4 steps was proposed. The generalization of the method as well as additional experiments will be communicated.

# 6    References

[1] P. Kranjnik, J. Kopac, Modern machining of die and mold tools, 2004, Journal of Materials and Processing Technology, 157–158: 543–552.
[2] B. Kecelj, J. Kopac, Z. Kampus, K. Kuzman, 2004, Speciality of HSC in manufacturing of forging dies, Journal of Materials Processing Technology, 157–158: 536–542.
[3] http://www.delcam.com
[4] http://www.mastercam.com
[5] S. Anderberg, T. Beno, L. Pejryd, 2009, CNC machining process planning productivity – a qualitative survey, In: Proceedings of The International 3'rd Swedish Production Symposium, SPS 09. - 978-91-633-6006-0 ; s. 228-235
[6] A. Flutter, J. Todd, 2001, A machining strategy for tool making, Computer-Aided Design, 33: 1009-1022.
[7] Rech J, Hamdi H, Valette S (2008) Workpiece surface integrity. In: Davim JP (ed) Machining: fundamentals and recent advances. Springer, London, pp 59–96
[8] C.K. Toh, 2005, Design, evaluation and optimisation of cutter path strategies when high speed machining hardened mould and die materials, Materials and Design, 26: 517–533.
[9] Axinte DA, Dewes RC, 2002, Surface integrity of hot work steel after high speed milling-experimental data and empirical models. J Mater Process Tech 127:325–335
[10] Kalvoda T, Hwang YR, 2009, Impact of various ball cutter tool positions on the surface integrity of low carbon steel. Mater Design 30:3360–3366
[11] M.C. Kang, K. K. Kim, D. W. Lee, J. S. Kim, N. K., Kim, 2001, Characteristics of inclined planes according to the variation of cutting direction in high-speed ball-end milling, International Journal of Advanced Manufacturing Technology, 17: 323–329.
[12] A.M. Ramos, C. Relvas, J.A. Simoes, 2003, The influence of finishing milling strategies on texture, roughness and dimensional deviations on the machining of complex surfaces, Journal of Materials Processing Technology, 136: 209–216.
[13] Y. B. Guo, W. Li, and I. S. Jawahir, 2009, Surface integrity characterizarion and prediction in machining of hardened and difficult-to-machine alloys: a state-of-art research review and analysis, Machining Science and Technology, 13: 437-470.
[14] N. Guillemot & B. K. Mawussi & C. Lartigue & R. Billardon, 2013, A first approach to characterize the surface integritygenerated by ball-end finishing milling, Int J Adv Manuf Technol, 64:269–279.
[15] L. Tapie, B.K. Mawussi, 2008, Decomposition of forging die for high speed machining, In IDMME -Virtual concept conference 2008, Beijing, China.
[16] K.B. Mawussi, L. Tapie, 2011, A knowledge base model for complex forging die machining, Computers & Industrial Engineering, 61: 84–97.

[17] S. Ding, M.A. Mannan, A.N. Poo, D.C.H. Yang, Z. Han, Adaptive Iso-planar tool path generation for machining of free-form surfaces, 2003, Computer-Aided Design, 35: 141-153.
[18] S. Ding, M.A. Mannan, A.N. Poo, D.C.H. Yang, Z. Han, The implementation of adaptive isoplanar tool path generation for the machining of free-form surfaces, 2005, International Journal of Advanced Manufacturing Technology, 26: 852–860.
[19] L. Tapie, B.K. Mawussi, B. Anselmetti, 2007, Circular tests for HSM machine tools: Bore machining application. International Journal of Machine Tools & Manufacture, 47: 805–819.
[20] L. Tapie, B.K. Mawussi, B. Anselmetti, 2007, Machining strategy choice: Performance viewer. In S. Tichkiewitch, M. Tollenaere, & P. Ray (Eds.), Advances in integrated design and manufacturing in mechanical engineering II, Dortech, Netherlands: Springer, 343-356.